\documentclass[11pt]{article}

\usepackage[utf8x]{inputenc}
\usepackage[T1]{fontenc}
\usepackage{amssymb,amsfonts,amsmath}
\usepackage{graphicx}
\usepackage{mathrsfs}
\usepackage{bm}
\usepackage{wasysym}
\usepackage{placeins}
\usepackage{lineno}
\usepackage{multirow}
\usepackage{times}
\usepackage{calc}
\usepackage{fullpage}

\begin{document}

\title{Exploring the isotopic niche: isotopic variance, physiological incorporation, and the temporal dynamics of foraging}
%
%
%

\author{
Justin D. Yeakel${}^{a,b,1,2}$, \\
Uttam Bhat${}^{a,c,2}$, \\
Emma A. Elliott Smith${}^{d}$, \\
Seth D. Newsome${}^{d}$, \\ \\
\small${}^a$Santa Fe Institute, Santa Fe, New Mexico, USA \\
\small${}^b$School of Natural Sciences, University of California, Merced, Merced, California, USA \\
\small${}^c$Department of Physics, Boston University, Boston, Massachusetts, USA \\
\small${}^d$Department of Biological Sciences, University of New Mexico, Albuquerque, New Mexico, USA\\
\small${}^1$To whom correspondence should be addressed. Email: jdyeakel@gmail.com \\
\small${}^2$J.D.Y. and U.B. contributed equally to this work.
}

\date{}

\maketitle

%
%
\section*{Abstract}

Consumer foraging behaviors are dynamic, changing in response to prey availability, seasonality, competition, and even the consumer's physiological state.
The isotopic composition of a consumer is a product of these factors as well as the isotopic `landscape' of its prey, i.e. the isotopic mixing space.
Stable isotope mixing models are used to back-calculate the most likely proportional contribution of a set of prey to a consumer's diet based on their respective isotopic distributions, however they are disconnected from ecological process.
Here we build a mechanistic framework that links the ecological and physiological processes of an individual consumer to the isotopic distribution that describes its diet, and ultimately to the isotopic composition of its own tissues, defined as its `isotopic niche'.
By coupling these processes, we systematically investigate under what conditions the isotopic niche of a consumer changes as a function of both the geometric properties of its mixing space and foraging strategies that may be static or dynamic over time.
Results of our derivations reveal general insight into the conditions impacting isotopic niche width as a function of consumer specialization on prey, as well as the consumer's ability to transition between diets over time.
We show analytically that moderate specialization on isotopically unique prey can serve to maximize a consumer's isotopic niche width, while temporally dynamic diets will tend to result in peak isotopic variance during dietary transitions.
We demonstrate the relevance of our theoretical findings by examining a marine system composed of nine invertebrate species commonly consumed by sea otters.
In general, our analytical framework highlights the complex interplay of mixing space geometry and consumer dietary behavior in driving expansion and contraction of the isotopic niche.
Because this approach is established on ecological mechanism, it is well-suited for enhancing the ecological interpretation, and uncovering the root causes, of observed isotopic data.


%

\section{Introduction}

Consumer foraging behaviors are dynamic, often resulting in variable diets that change over space and time as a function of environmental conditions, the densities of consumers and available resources, and even the physiological states of individual foragers, to name a few.
Understanding how diets change, and to what extent different conditions promote or inhibit specific changes, is both a challenging theoretical and empirical problem in ecology, but is essential for elucidating the adaptive nature of complex ecological systems.

The comparison of carbon ($\delta^{13}\rm{C}$) and nitrogen ($\delta^{15}\rm{N}$) stable isotope values of a consumer with respect to its potential prey is a commonly utilized approach to quantify diet composition.
The carbon and nitrogen isotopic composition of a consumer represents that of the food it eats, but is offset by predictable amounts, often called trophic discrimination factors, that are mediated by consumer physiology \cite{VanderZanden:2001p2449,Post:2002wn,FoxDobbs:2007kx,Bearhop:2010db}.
Assuming that trophic discrimination factors have been accounted for, the isotopic composition of a consumer thus reflects
1) the proportional contribution of different prey to the consumer's diet \cite{Moore:2008kg,Parnell:2010ub}, and
2) the isotopic composition of its prey, collectively described as the isotopic, or prey, mixing space \cite{Phillips:2001ws,Phillips:2002tu,Newsome:2007tz}.

The isotopic niche of a consumer is a low-dimensional specification of the `Hutchinsonian niche' \cite{Hutchinson:1957tg}, an n-dimensional hypervolume that defines all biotic and abiotic requirements needed for a species to exist.
The isotopic niche is also generally derived from both biotic and abiotic processes, but in contrast to the large and ultimately immeasurable construct of the n-dimensional hypervolume, isotopic niches are defined exactly as a consumer's isotopic variance with a dimension determined by the number of isotope systems employed \cite{Bearhop:2004im,Newsome:2007tz,Newsome:WhhVfocb,delRio:2011p2553,Jackson:2011kd}.
It is the width, or spread of this distribution that correlates to the breadth of the isotopic niche.
Although the isotopic niche can be the result of many ecological and environmental factors that influence the flow of elements through biological systems \cite{Araujo:2007iua}, when it is primarily driven by consumer-resource interactions, the isotopic niche is synonymous with the trophic niche as defined by \cite{Bearhop:2004im}.
Changes in the size of the isotopic niche have been shown to relate to ecosystem fragmentation \cite{Layman:2007ky}, evolutionary diversification \cite{Fedosov:2014hm}, changes in food availability \cite{Lehmann:2015db}, and even individual responses to seasonal environments \cite{delRio:2011p2553}, however a systematic understanding of how different sources of variability influence the isotopic niche is lacking.

Both the consumer's dietary strategy as well as the geometry of the isotopic mixing space (defined collectively as the isotopic distributions of potential prey available to the consumer) contribute different sources of variability that are reflected in the isotopic composition of the consumer.
For example, a consumer could be an obligate specialist on a single prey (low dietary variability), in which case the isotopic mean and variability of the consumer's tissues will reflect that of its targeted prey.
In contrast, the consumer could be a generalist, or intermediate specialist on certain prey such that the variability in its diet as well as the isotopic variability of its prey contributes to its own isotopic composition.
Thus, the isotopic composition of a consumer is not only a reflection of what the consumer eats, but is also determined by the mixing space over which it incorporates isotope values, and the amount of time over which a particular tissue integrates dietary inputs.

The isotopic niche is generally defined with respect to the isotopic variance of a population \cite{Araujo:2007iua,Araujo:2009p2286,Fink:2012eg}, however population-level variance is itself an artifact of the individual-level distributions from which the population is composed \cite{Bolnick:2007p1253,Araujo:2011gm}.
An important step in understanding how the isotopic niche changes in response to different consumer foraging strategies is to examine how individual-level variation contributions to observed isotopic variability.
Individual-level variability, in particular that variability resulting from temporal changes in diet, may have large implications for the population \cite{Schreiber:2011wx,Gibert:tz} and by extension the community \cite{KOKKORIS:2002tf,Kondoh:2003p1625,Yeakel:OfhI8s6v}.
For example, populations with greater individual-level variability have been shown to exhibit greater stability \cite{Bolnick:2011gj,Gilbert:2014ee}, and in certain cases can increase the likelihood of species coexistence \cite{Schreiber:2011wx,Vasseur:2011ek} and competitor persistence \cite{Gibert:2015kc}.
Changes in individual diet through time are frequently measured via stable isotope analysis of metabolically inert, temporally-integrating tissues such as hair, whiskers, claws, or even growth layers in teeth \cite{Koch:1995vj,Matthews:2004hw,Sponheimer:2006fj,Post:2008ki,Newsome:2009tn,Yeakel:2009hz,Hopkins:2015ip}.

Once physiologically-mediated isotopic discrimination between a consumer and its food has been accounted for, the isotopic niche is a direct reflection of the proportional contribution of different prey resources to the consumer's diet.
Isotope mixing models, which originally used a bootstrapping framework \cite{Phillips:2003kq,Phillips:2005p1007} and now employ Bayesian methods \cite{Moore:2008kg,Parnell:2010ub,Hopkins:2012dza,Parnell:2012wv}, are used to reconstruct the probability distribution that describes the contribution of different prey. 
These tools are forensic in nature, and can be used to back-calculate consumer diets across a range of isotopically distinct prey with differing stoichiometries \cite{Hopkins:2015ip}, to quantify intra- and inter-population niche variability \cite{Semmens:2009uq}, to reconstruct diets of extinct taxa \cite{Yeakel:2012uc}, and even combined with process-based models that are designed to constrain contribution-to-diet estimates based on known mechanistic relationships between species \cite{Ogle:2014jg}.

A conceptually divergent, yet parallel, strategy is to consider the inter-related effects of ecological interactions and the concomitant integration of stable isotope values, and how these factors combine to result in the isotopic composition of a consumer.
In this way, the consumer's isotopic distribution can be forward-integrated through time based on the mechanistic linkages between its foraging strategy, its ability to find and acquire prey, and the isotopic mixing space over which its diet is composed.
This general perspective has been applied to investigate properties of the isotopic niche for cases of fixed (non-varying) diets \cite{Araujo:2007iua,Araujo:2009p2286}, and with respect to experimental systems where consumers are forced to switch between unique prey \cite{Fink:2012eg}.
However, such a process-based framework has not been directly linked to consumer foraging behaviors, prey availability, or other ecological considerations such that the combined impacts of ecologically-mediated dietary variation and mixing space geometry on the isotopic niche can be assessed.


Here we build a mechanistic framework that links the ecological and physiological processes of an individual consumer to the isotopic distribution that describes its diet, and ultimately to the isotopic composition of its tissues.
By coupling these processes, we systematically investigate under what conditions the isotopic niche of a consumer changes as a function of both alternative foraging strategies, and geometric properties of its mixing space.
We show that the isotopic variance of a consumer's diet (and by extension the width of its isotopic niche) is expected to be a concave parabolic function with respect to prey specialization, such that it possesses a peak variance, though the appearance of this peak is contingent on mixing space geometry.
We demonstrate the potential importance of these findings by examining a prey mixing space for sea otters (\emph{Enhydra lutris}), which are known to possess high dietary individuality \cite{Estes:2003uc,Tinker:2008tp,Newsome:2009tn,Newsome:2015gy}.

We then extend our framework to explore how a consumer's isotopic niche responds to temporally dynamic diets.
We examine an example of a switch between two probabilistic diets that oscillates between seasons, and show analytically that the maximum expected isotopic niche width occurs during dietary transitions.
Determining how different sources of variability emerge from ecological, behavioral, and geometric drivers is important for interpreting the meaning of concepts such as the isotopic niche.
Moreover, integrating these different sources of variation into a statistical process-based framework lends itself particularly well to hypothesis-testing aimed at uncovering the root causes of observed isotopic data.
We hope that our approach is a step towards assessing how the isotopic niche may change due to more complex foraging behaviors and even population dynamics, and we expand on these potential avenues in the Discussion.

%





\section{Methods \& Analysis}
We begin by establishing a forward-integration approach for modeling the incorporation of stable isotopes from multiple resources into a consumer's tissues.
This new methodology provides an analytical link between the mechanistic drivers of foraging and the distribution of stable isotope values describing a consumer's tissues over time.
Using this framework, we aim to
1) examine how certain dietary behaviors, such as prey specialization and different modes of dietary variation, impact the isotopic variance of consumer tissues irrespective of the statistical model used to formulate consumer diets, and
2) show how these methods can be expanded to include foraging behaviors that themselves are temporally dynamic, changing over seasons or years.
Accordingly, our general goal is to reveal how both ecological and geometric factors can influence the expansion and contraction of isotopic variability, thus aiding ecological interpretation of the `isotopic niche'.

\subsection*{Deriving the within-individual isotopic niche width}
There are many ways to statistically summarize the integration of prey by a consumer species, however in order to establish a mechanistic link between foraging and the consumer's isotopic composition, we follow the proceeding heuristic foraging mechanic.
We assume that a consumer encounters and consumes resources in proportion to the encounter rate of each prey; prey that are encountered more frequently are assumed to be consumed more frequently.
An alternative approach could incorporate preferences \cite{Chesson:1983bt} or even state-dependence \cite{Mangel:1988uaa,Clark:2000tra,Mangel:2014kz}, and we will briefly address these considerations in the Discussion.
As prey are encountered and consumed, the prey's isotope values are incorporated into the consumer's tissues weighted by the prey-specific proportional contribution to diet.
The resulting distribution that describes the dietary input of multiple prey (each with  isotope values that are independently and Normally distributed) is a mixed Normal distribution with weights determined by the prey's proportional contribution to diet.
This proportional contribution is itself a random variable drawn from a Dirichlet density (a multivariate Beta distribution) that serves as a probabilistic description of the consumer's dietary input \cite{Ainsworth:2010uo}.

The following section details our probabilistic determination of the consumer's isotopic composition.
We focus our attention on the variability of the isotopic distribution describing the consumer's diet, which scales directly with the consumer's own isotopic distribution, itself equivalent to the isotopic niche \cite{Bearhop:2004im,Newsome:2007tz}.
Here and henceforth, we assume that the isotope ratios under consideration follow dietary pathways such that the isotopic niche is synonymous with the trophic niche \cite{Bearhop:2004im}.

A consumer encounters each prey at a frequency determined by a Poisson process with parameter $\psi_i$, which determines the number of encounters $M_i=m_i$ between time 0 and time $t$.
Allowing only the encounter rate to vary, the maximum entropy process (or the process that assumes no other structure) is the Poisson process, such that $m_i$ varies according to the frequency distribution

\begin{equation}
f_{M_i} (m_i|\psi_i) = {\rm e}^{-\psi_i t}\frac{(\psi_i t)^{m_i}}{m_i!}.
\end{equation}

\noindent Here and henceforth, we use the general function $f(\cdot)$ to denote different frequency distributions, as well as uppercase notation to describe stochastic variables, and lowercase notation to describe specific values of stochastic variables.
If we assume that encounter rates $\psi_i$ are themselves variable, such that some prey are more patchily distributed than others, we can treat $\Psi_i = \psi_i$ as a random variable with a Gamma density

\begin{equation}
f_{\Psi_i} (\psi_i | c, a_i) = \frac{c^{a_i}}{\Gamma (a_i)}{\rm e}^{-c \psi_i}\psi_i^{a_i - 1}.
\end{equation}

\noindent \noindent where $\Gamma(\cdot)$ is the gamma function \cite[cf.][]{Mangel:2006wa}. 
Here, $a_i$ is the dispersion parameter, which is proportional to the mean encounter rate, and $c$ measures the time between encounters \cite{Mangel:2006wa,Ainsworth:2010uo,Yeakel:2013hi}.
Integrating across all possible values of $\psi_i$, we obtain the Negative Binomial density with mean encounter rate $a_i/c$ and coefficient of variation $1/\sqrt{a_i}$ \cite{Hilborn:1997ds}.
Following the derivation described by \cite{Ainsworth:2010uo}, if we define the proportional contribution of prey to a consumer's diet to scale with the encounter rate, such that

\begin{equation}
\label{eqp}
  p_i = \frac{\psi_i}{\sum_{j=1}^n \psi_j},
\end{equation}

\noindent then the random variable $P_i = p_i$ where $P_i \in {\bm P}$ and $p_i \in {\bm p}$, and $\sum_i p_i = 1$, where boldface type denotes vectors of variables.
From Eq. \ref{eqp}, it follows that the vector describing the proportional contributions to diet $\bm P$ follows a Dirichlet distribution \cite{Johnson:1960gu} with density

\begin{equation}
  f_{\bm P}(p_1,...,p_n|a_1,...,a_n) = \frac{\Gamma(\sum_{i=1}^n a_i)}{\sum_{i=1}^n\Gamma(a_i)}\prod_{i=1}^n p_i^{a_i - 1}.
\end{equation}

\noindent As such, the expected proportional contribution of a prey $i$ to the consumer's diet has the expectation ${\rm E}\{P_i\}=a_i/a_0$ where $a_0 = \sum_i a_i$, and variance

\begin{equation}
  \label{eqDirVar}
  {\rm V}\{P_i\} = \frac{a_i(a_0 - a_i)}{a_0^2(a_0 + 1)}.
\end{equation}

\noindent In this paper we consider only the case where resources are plentiful and feeding is regular (such that consumers do not starve), and we draw a single prey $i$ with probability $p_i$ for inclusion to the consumer's diet.

Describing the dietary behavior of a consumer as a Dirichlet distribution provides a flexible and powerful framework to investigate how different foraging strategies influence a consumer's isotopic niche.
For example, a pure generalist consumer would have a Dirichlet distribution with parameters $a_i = 1$ for all prey $i=1,...,n$, such that the marginal distribution for $P_i$ is close to uniform with expectation ${\rm E}\{P_i\} = 1/n$.
Because we have assumed that the proportional contribution of a prey to the consumer's diet scales with the prey's encounter rate, this would be analogous to a system where a consumer is equally likely to encounter the same number of any prey.
In contrast, an obligate specialist would have a Dirichlet density that is spiked for a given prey $k$, where the single parameter $a_k \gg 1$, while $a_{i \neq k} = 1$.
The use of a Dirichlet distribution is also at the heart of Bayesian isotope mixing models \cite{Moore:2008kg,Parnell:2010ub,Hopkins:2012dza,Parnell:2012wv}, which assume a Dirichlet prior and enable the input of alternative dietary information to inform isotopic data.

If the stable isotope ratios for each of the potential prey follow independent Normal distributions, and the dietary behavior of the consumer has a Dirichlet density, the resultant density that describes the isotopic distribution of a consumer's diet, $f_Z(z){\rm dz} = {\rm Pr}(z \leq Z \leq z + {\rm dz}) + o({\rm dz})$, is a mixed Normal distribution with weights given by $\bm p$ drawn from the Dirichlet.
Given that the isotopic means and variances for prey $i$ are denoted by $\mu_i$ and $\sigma^2_i$, respectively, this density can be written as


\begin{equation}
  \label{eqfZ}
f_Z(z|{\bm a},{\bm \mu},{\bm \sigma}) = \left(\sum_{i=1}^n p_i\frac{1}{\sqrt{2 \pi \sigma_i^2}}{\rm e}^{-\frac{(z-\mu_i)^2}{2\sigma_i^2}}\right)f_{\bm P}(\bm p | \bm a),
\end{equation}

\noindent with the expectation

\begin{equation}
\label{eqEZ}
  {\rm E}\{Z\} = \sum_{i=1}^n \frac{a_i}{a_0} \mu_i,
\end{equation}

\noindent Accordingly, the isotopic mean of the consumer's diet is a weighted average of the isotopic means of its potential prey, where weights are determined by the outcome of the Dirichlet random variable.

Of more interest to us here is the variance of $Z$, which will allow us to analytically determine the isotopic niche width of the consumer as a function of its dietary behavior and the mixing space of its prey.
We find that

\begin{equation}
\label{eqVarZ}
  {\rm V}\{Z\} = \sum_{i=1}^n \frac{a_i}{a_0}\left(\sigma_i^2 + \mu_i^2\right) - \frac{a_i^2\mu_i^2}{a_0^2}-\sum_{i \neq j}\frac{a_i a_j \mu_i \mu_j}{a_0^2}.
\end{equation}

\noindent Although the form of Eq. \ref{eqVarZ} is not intuitive, we emphasize that - over different dietary behaviors that shape the Dirichlet distribution and for different isotopic mixing space geometries - it is this equation that governs the expansion or contraction of the consumer's isotopic niche width, and therefore of chief ecological interest.

The isotopic variance of the consumer's diet ${\rm V}\{Z\}$ can be simplified by considering a specific set of dietary behaviors.
Here we examine how ${\rm V}\{Z\}$ is influenced by generalist vs. specialist consumer diets, as well as the role of mixing space geometry, in determining consumer isotopic niche width.
It is important to note that specialism, as discussed here, defines the degree to which a consumer's diet is dependent on a single prey resource, ranging from $s_k=1/n$ (obligate generalization on prey $k$) to $s_k=1$ (obligate specialization on prey $k$).
It is thus conceptually different than `individual specialization' defined as the proportional variance of an individual relative to that of its population \cite[Within-Individual Component / Total Niche Width, or WIC/TNW;][]{J:1979wc}, and is often the variable of interest in other studies examining properties of the isotopic niche \cite{Araujo:2007iua,Araujo:2009p2286,Araujo:2011gm,Layman:2011cm}.

If a generalist consumer alters its diet to include more of a certain prey $k$ relative to the others, the Dirichlet distribution that defines its dietary behavior changes from $a_i=1$ for all $i=1,...,n$ to $a_{i \neq k}=1$ for $i=1,...,n$, with $a_k>1$.
As specialization increases, the Dirichlet parameter corresponding to the targeted prey $k$ increases to a value much higher than one (pure specialization is obtained only at the limit $a_k \to \infty$).
Thus, we can assume that $a_i=1$ for all $i \neq k$, and $a_k = (n-1)s_k/(1-s_k)$, where $s_k$ denotes specialization on prey $k$. 
We can thus substitute $a_0 = (n-1)/(1-s_k)$ and $p_i = a_i/a_0 = (1-s_k)/(n-1)$ for all $i \neq k$, and $a_k/a_0 = s_k$, allowing us to rewrite Eq. \ref{eqVarZ} in terms of $s_k$, such that

\begin{equation}
\label{eqVarZs}
{\rm V}\{Z\} = \frac{1-s_k}{n-1}\sum_{i \neq k}^n \left(\sigma_i^2 + \mu_i^2\right) + s_k(\sigma_k^2 + \mu_k^2) - \left(\frac{1-s_k}{n-1}\sum_{i \neq k}^n \mu_i + s_k\mu_k \right)^2,
\end{equation}

\noindent and note that, independent of the prey mixing space (a function of $\mu_i$ and $\sigma_i^2$ for prey $i=1,...,n$), the isotopic variance of the consumer's diet will always be a concave parabolic function over $s_k$.
With respect to the size of the consumer's isotopic niche width, this means that there can be a peak variance for a value of $s_k$ intermediate to pure generalization ($s_k=1/n$) and pure specialization ($s_k=1$).

The peak variance $\hat s_k$, which describes the maximum isotopic variance of the consumer, may or may not fall between $s_k=1/n$ and $s=1$, and is only of ecological interest if it does.
The peak variance can be solved analytically by setting the derivative of Eq. \ref{eqVarZs} with respect to $s_k$ equal to zero, which results in

\begin{equation}
	\hat s_k = \frac{A(1-n)+B (n-1)^2+2 C (C-D n+D)}{2 (C-D n+D)^2},
\end{equation}

\noindent where $A = \sum_{i \neq k}^n \left(\sigma_i^2 + \mu_i^2\right)$, $B = \left(\sigma_k^2 + \mu_k^2\right)$, $C = \sum_{i \neq k}^n \mu_i$, and $D = \mu_k$.

Determination of the peak variance allows us to predict where the consumer's isotopic niche is expected to be maximized as a function of specialization on different prey.
Importantly, we note that the concave parabolic nature of ${\rm V}\{Z\}$ is a property of the mixed normal distribution that characterizes the geometry of the prey mixing space, and not the underlying statistical model used to formulate consumer foraging behavior.
This can be seen by observing that the quadratic term in Eq. \ref{eqVarZ} ($-a_i^2\mu_i^2/a_0^2$ where $a_i/a_0$ is the mean of the proportional contribution of prey $p_i$, which weights the influence of the different prey isotope distributions in $f_Z$) appears with a negative sign.
Thus, regardless of the process that leads to $p_i$, the term will always be concave parabolic over prey specialization, such that this result will apply to any underlying foraging model.

Although here we have focused on the special case where a consumer targets a single prey, one can rewrite the equation for the consumer's isotopic niche width with respect to increasing specialization on any number or combination of prey in the mixing space.
For example, in the case where a consumer specializes on two prey (e.g. two species of crab), one would rewrite Eq. \ref{eqVarZ} in terms of both $s_k$ (specialization on prey $k$) and $s_l$ (specialization on prey $l$), resulting in a concave parabolic plane in dimensions $s_k$ and $s_l$.
Determining the maximum variance would then entail taking the derivative of Eq. \ref{eqVarZ} with respect to both $s_k$ and $s_l$.
In dimensions higher than 2, the process would be the same, with the goal of finding the maximum variance over a hyperplane with a number of dimensions determined by the number of prey on which the consumer is preferentially targeting.
Because specializing on multiple prey does not introduce anything conceptually unique, we consider only the case of a single-prey specialist.

\subsection*{The Dynamics of Isotopic Incorporation}
We have established a framework for calculating analytically the distribution of isotope values that characterizes a consumer's diet, composed of multiple, isotopically distinct prey.
The dietary behavior of the consumer is a function of a single Dirichlet distribution, which is assumed not to change over time, although we will relax this assumption in the next section.
Over long timescales the dietary distribution of the consumer is static, with a fixed mean and variance.
Over short timescales, the consumer's diet varies as Eq. \ref{eqDirVar}, while its final isotopic distribution has a variability emerging from the combined effects of the Dirichlet and the mixed Normal distribution describing the prey mixing space (Eq. \ref{eqVarZ}).

As the consumer acquires and consumes its prey, the isotopic composition of its diet is incorporated into its tissues.
The timescale of physiological incorporation is based on the turnover rate of consumer tissues, which on the fast end can occur within days to weeks (e.g. blood plasma), and on the slow end occur over years (e.g. bone) \cite{Tieszen:1983ij}, and can be estimated via controlled feeding studies \cite{Kurle:2009ch,Bearhop:2010db,Kim:2012kc}. 
Although the physiological details are not well understood, isotopic incorporation can be modeled using either single- or multi-compartmental approaches \cite{Cerling:2006kv,delRio:2008bs}.
In a single compartment framework, isotope ratios are ingested with food, and directly incorporated into consumer tissues at a tissue-specific rate.
In multiple compartment frameworks, it is assumed that incorporation occurs over multiple body pools, the turnover of each potentially occurring at different rates.
More recent approaches incorporate specific metabolic pathways to model the flux of stable isotopes within body tissues \cite{Pecquerie:2010kp}.

In this next section, we assume that the ingested isotope ratios are incorporated into consumer body tissues directly, moderated by the rate of incorporation $\lambda$, which is treated as a free parameter.
Here we consider only a single compartment model, such that isotope ratios are directly shuttled to consumer tissues at rate $\lambda$; we note, however, that functions for multi-compartment models could be used instead, though we do not expect large qualitative differences in results \cite[cf. Fig. 1 in][]{delRio:2008bs}.
For simplicity, we assume that time is scaled such that a single time step corresponds to a single foraging bout.
Moreover, we assume that the consumer is incorporating prey of smaller size than itself, such that $ 0 < \lambda < 1$.
Thus, we aim to determine the isotopic composition of the consumer $X_c$ as a function of its diet, mixing space geometry, and $\lambda$.
We note 1) that the isotopic composition of the consumer could represent its carbon ($\delta^{13}{\rm C}$) or nitrogen ($\delta^{15}{\rm N}$) isotope distribution, and our proceeding derivations work equivalently for both, and 2) that all trophic discrimination factors are assumed to have been accounted for, such that $X_c$ directly reflects the consumer's diet.
%
%
Taking into account the stochastic effects described in the previous section, including the variation associated with the consumer's diet and the isotopic variation associated with each prey, we describe changes in the consumer's isotopic distribution with the stochastic differential equation

\begin{equation}
\label{eqSDE}
{\rm d}X_c(t) = \lambda\left({\rm E}\{Z\}{\rm dt} + \sqrt{{\rm V}\{Z\}}{\rm dW}\right) - \lambda X_c(t){\rm dt}.
\end{equation}

\noindent where ${\rm dW}$ is the increment of Brownian motion.
This stochastic differential equation describes an Ornstein-Uhlenbeck process, which is a stochastic process that has a steady state variance around the mean \cite{Mangel:2006wa}.
Because the time interval ${\rm dt}$ is infinitesimal at the continuous limit, the consumer's isotopic distribution will have a Normal distribution.
In this case, if the initial isotopic values of the consumer at time $t=0$ is $X_c(0)$, the expectation and variability of $X_c$ at time $t$ are

\begin{align}
  \begin{split}
    \label{eqEVar}
{\rm E}\{X_c(t)\} &= {\rm E}\{Z\} + (X_c(0) - {\rm E}\{Z\}){\rm e}^{-\lambda t},\\
{\rm V}\{X_c(t)\} &= \frac{\lambda {\rm V}\{Z\}}{2}\left(1 - {\rm e}^{-2\lambda t}\right).
\end{split}
\end{align}

\noindent where ${\rm E}\{Z\}$ and ${\rm V}\{Z\}$ are as defined in Eqns. \ref{eqEZ} and \ref{eqVarZ}.
One can observe that as $t$ increases, the exponential part of ${\rm E}\{X_c(t)\}$ and ${\rm V}\{X_c(t)\}$ go to zero, such that ${\rm E}\{X_c(t)\} \to {\rm E}\{Z\}$, and ${\rm V}\{X_c(t)\} \to \lambda{\rm V}\{Z\}/2$.
In other words, the expectation of the consumer's isotopic distribution will equilibrate to that of its diet, while its variance will always be less than the variance of its diet by a factor of $\lambda/2$.
Variance decreases as the rate of incorporation decreases due to the consumer averaging its isotopic value over more prey (because the tissue is turning over more slowly), and this serves to average out fluctuations in the consumer's diet.



Our static model is defined by a consumer's diet that varies instantaneously over a given parameterization of $f_Z(z)$.
This is relevant for organisms that have a consistently varying diet over time, however most organisms have diets that undergo large, qualitative changes over longer periods time.
In such cases, the Dirichlet distribution that characterizes diet during one small temporal interval will be different than the Dirichlet distribution characterizing diet during another interval far apart in time.
Such a shift might be due to seasonal, ontogenetic, or demographic changes in the consumer or its prey base over the course of months, or years, depending on the timescale of interest.
In the following section, we will relax the assumption that diet is characterized by a single Dirichlet distribution, thus generalizing our formulation of consumer isotopic dynamics as a function of time.

The random variable of interest is now $Z(t)$, which is the trajectory defining the isotopic distribution of the consumer's diet over time.
Solving for $X_c(t)$, we find

\begin{align}
  \label{eqEVarZt}
{\rm E}\{X_c(t)\} = X_c(0){\rm e}^{-\lambda t} + \lambda{\rm e}^{-\lambda t} \int_{s=0}^t {\rm e}^{\lambda s} {\rm E}\{Z(s)\}{\rm d}s, \nonumber \\
{\rm V}\{X_c(t)\} = \lambda^2 {\rm e}^{-2\lambda t} \int_{s=0}^t {\rm e}^{2\lambda s} {\rm V}\{Z(s)\} {\rm d}s.
\end{align}

\noindent By defining the temporal dynamics of diet $Z(t)$ and the incorporation rate $\lambda$, we can thus analytically determine the isotopic mean and variance of the consumer's tissues.



\section{Results}

\begin{figure}[h!]
\begin{center}
\includegraphics[width=0.75\textwidth]{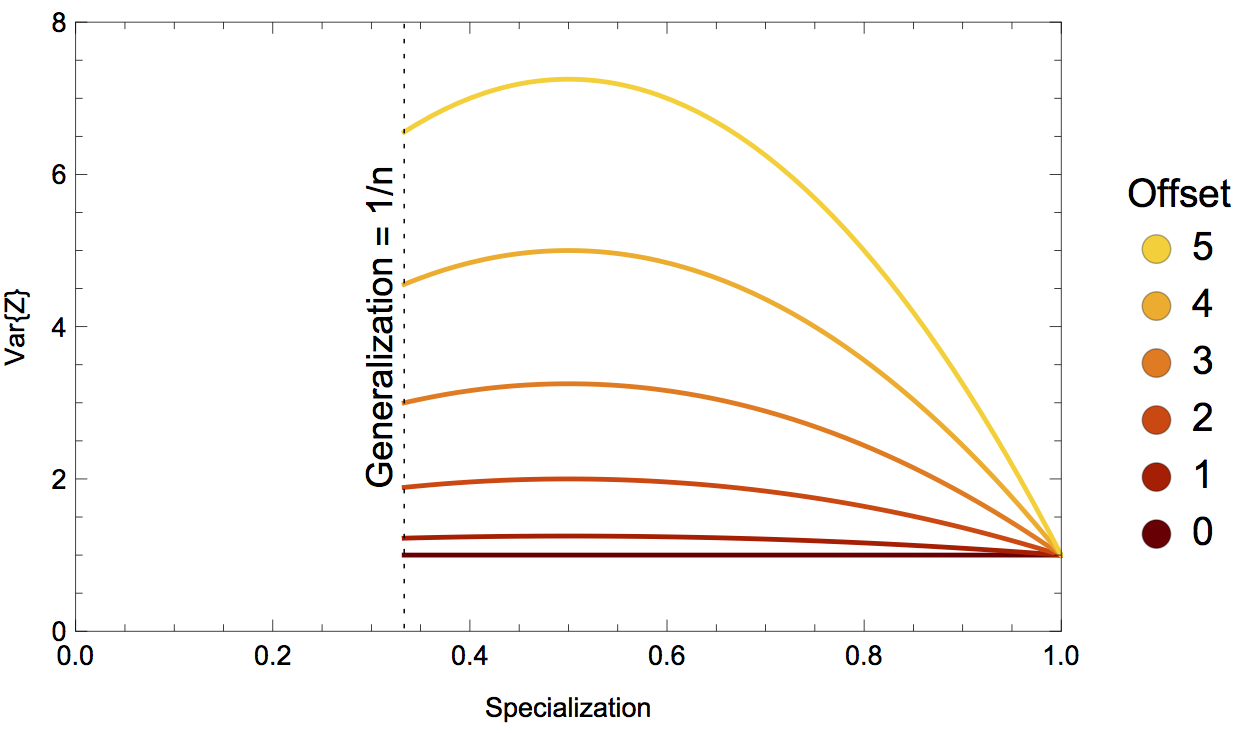}
\end{center}
\textbf{\refstepcounter{figure}\label{figvar} Figure \arabic{figure}.}{
Variance of the isotopic distribution of a consumer's diet, ${\rm V}\{Z\}$, with respect to specialization on a single prey, $s_k$.
This illustrative example shows a three-prey system with prey means $\{-15,-15+\mbox{offset},-15\}$ and equal variances; colors depect specialization on prey 2 with a mean isotope value that is a function of some offset amount.
As the offset of the targeted prey increases, so does the nonlinearity of ${\rm V}\{Z\}$.
}
\end{figure}

We have provided an analytical solution for the mean and variance of the consumer's isotopic distribution as a function of its diet and prey mixing space.
By formulating these solutions in terms of consumer generalization and specialization (Eq. \ref{eqVarZs}), we make three observations:
1) the variance of the isotopic distribution of the consumer's diet, ${\rm V}\{Z\}$, which scales to its isotopic niche width, is concave parabolic (Fig. \ref{figvar}); 
2) whether and to what extent ${\rm V}\{Z\}$ demonstrates measurable nonlinearity depends in part on the geometry of the mixing space;
3) the peak variance over the generalization-specialization continuum is the consumer's maximum isotopic niche width.
This point may or may not exist at a value intermediate to an obligate generalist and obligate specialist.

The nonlinear nature of the consumer's isotopic niche width as a function of its specialization on certain prey (or combinations of prey) is driven almost entirely by the geometry of the prey mixing space.
One can gain some intuitive understanding of this nonlinearity by considering the following example, illustrated in Fig. \ref{figvar}.
In a three-prey system, where all prey have equal isotopic means and variances, a consumer that ranges from generalization on all three prey to specialization on a single prey will likewise have isotopically equivalent diets.
As the mean isotope value of the targeted prey is moved away from the others, such that its offset from the mixing space centroid \cite[the center of the mixing space;][]{Layman:2007vi,Newsome:WhhVfocb} is increased, the variance function displays increasing nonlinearity.

For a skewed mixing space, where one prey source has a very different isotope composition than the rest \cite[e.g. a mixing space consisting of terrestrial foods vs. a marine subsidy;][]{Newsome:2004p992}, a consumer incorporating isotopes from all three sources in equal proportions (a generalist) will have relatively higher isotopic variance than if its prey exhibited a less skewed mixing space geometry.
The skewness of the mixing space increases with the offset of the targeted prey from the mixing space centroid as shown in Fig. \ref{figvar}.
As the consumer integrates this isotopically unique prey in greater proportions, the heterogeneity of incorporated isotope values will increase, serving to increase the consumer's isotopic variability.
The isotopic variability will then decline as the consumer begins specializing on the atypical prey, and if it is consuming this prey exclusively, the isotopic variability of its diet will reflect the isotopic variability of its prey exactly.
The concave parabolic nature of the isotopic variability of the consumer's diet can thus be explained by heterogeneous incorporation of isotope ratios over an skewed, or asymmetric, mixing space.

Understanding what dietary strategy or mixing space geometry can maximize the isotopic niche width of the consumer's diet will serve to help ecologists determine what mechanisms - ecological or statistical - may be driving patterns in isotope data, or whether these mechanisms can be decoupled at all.
Our analytical solution for peak variance over dietary specialization on prey $k$, $\hat s_k$, reveals that maximum isotopic niche width can, but doesn't always, fall in $s_k \in [1/n,1]$, with bounds denoting exclusive prey generalization or specialization, respectively.
If the peak lies outside of this region, changes in isotopic variance as specialization on a targeted prey is increased will appear monotonic or even linear.

\begin{figure}[h!]
\begin{center}
\includegraphics[width=1\textwidth]{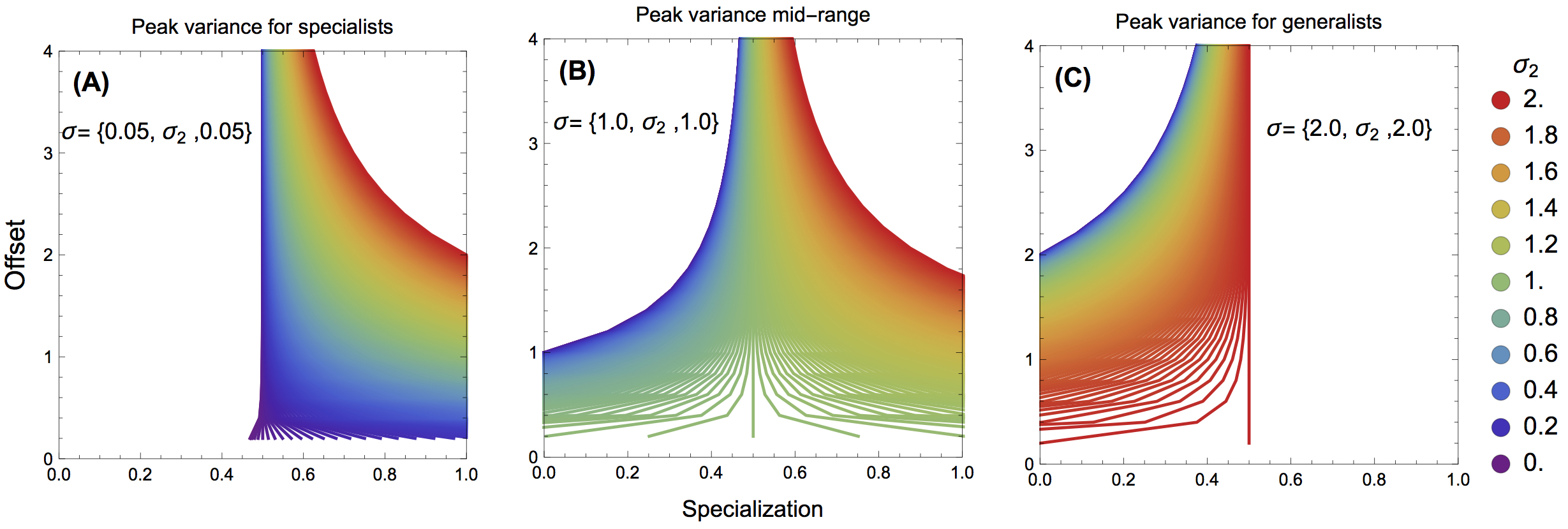}
\end{center}
\textbf{\refstepcounter{figure}\label{figspecvar} Figure \arabic{figure}.}{
Maximal consumer isotopic variance $\hat s_k$ over the specialization index $s$ as a function of mixing space geometry. 
{\bf(A)} and {\bf (B)} If the targeted prey has a higher than average isotopic variance, the maximum consumer niche width will lie towards consumer specialization.
{\bf (B)} and {\bf(C)} If the targeted prey has a lower than average isotopic variance, the maximum consumer niche width will like towards consumer generalization.
{\bf(A)}, {\bf (B)}, and {\bf(C)} as the mean offset of the targeted prey is farther from the centroid of the mixing space, the maximal consumer isotopic niche width tends towards $s=0.5$.
}
\end{figure}

Although the specific nature of $\hat s_k$ will depend strongly on mixing space geometry, we can elucidate certain key attributes that will determine the general nature of where this value falls.
For mixing space geometries where the targeted prey has higher than average variance, $\hat s_k$ will tend to lie towards prey specialization ($s_k>0.5$), however the offset of the mean value of the targeted prey from the mixing space centroid will quickly push $\hat s_k$ to $s_k \to 0.5$ (Fig. \ref{figspecvar}A,B).
In contrast, if the targeted prey has lower than average variance, $\hat s_k$ will tend to lie towards prey generalization ($s_k < 0.5$; Fig. \ref{figspecvar}B,C).
As before, if the offset of the targeted prey's mean value increases, $\hat s_k \to 0.5$.
In both cases, if the mean value for the targeted prey is close to the mixing space centroid, the maximum isotopic variance for the consumer could lie in any region.

\subsection*{Temporally variable diets}
The equilibrial solution to our stochastic differential equation (Eq. \ref{eqEVar}) reveals that the isotopic variability of the consumer scales to diet as a factor of $\lambda/2$.
As the incorporation rate decreases, such that the turnover time is longer, the isotopic variability of the consumer declines.
This is due to the consumer averaging its tissues over a greater number of foraging bouts.
Moreover, we observe that as the consumer transitions from some initial isotopic state $X_c(0)$ to diet, the variance of the consumer's isotopic values equilibrate twice as fast as the mean value.

If the consumer's diet is itself variable over time, we do not expect its isotopic composition to equilibrate as it would in a controlled feeding study (Eq. \ref{eqEVarZt}).
For example, the consumer might adopt one diet during the wet season, and another during the dry season, such that it oscillates between the two throughout the year.
We consider a composite diet with an isotopic distribution $\mathbb{Z}(t) \sim f_{\mathbb{Z}(t)}({\mathbb z}(t))$ that dynamically oscillates between two subdiets, which we will refer to as `seasonal diets' with frequency $\omega$.
We note that $1/\omega$ in this context corresponds to the `dietary correlation time' of \cite{Fink:2012eg}.
Seasonal diets have random variables $Z_1$ and $Z_2$, each distributed according to Eq. \ref{eqfZ}, though they have different underlying Dirichlet distributions -- encoding which prey the consumer targets during each season with frequency distributions $f_{\bm P_1}$ and $f_{\bm P_2}$ -- while the isotopic distributions of prey are assumed to be constant.
We can thus describe the composite diet as a mix of the seasonal diets characterized by weights that oscillate over time, and this determines the contribution of each seasonal dietary strategy to the whole.
We define $\mathcal{U}(t)$ to be the proportional contribution of $Z_1$ to the composite diet $\mathbb{Z}(t)$ over time, such that it can vary between zero (no incorporation of $Z_1$) to unity (complete reliance on $Z_1$).
The frequency distribution for the composite diet is thus


\begin{equation}
  f_{\mathbb{Z}(t)}(\mathbb{z}(t)) = \mathcal{U}(t) f_{Z_1}(z_1) + (1-\mathcal{U}(t)) f_{Z_2}(z_2). 
\end{equation}


If we do not specify the type of oscillation that drives changes in diet over time, the expectation and variance for the isotopic distribution of the composite diet over time are thus

\begin{align}
  \label{eqZtgen}
  \begin{split}
    {\rm E}\{\mathbb{Z}(t)\} &= \mathcal{U}(t){\rm E}\{Z_1\} + (1-\mathcal{U}(t)){\rm E}\{Z_2\}, \\
    {\rm V}\{\mathbb{Z}(t)\} &= \mathcal{U}(t){\rm V}\{Z_1\} + (1-\mathcal{U}(t)){\rm V}\{Z_2\} + \mathcal{U}(t)(1-\mathcal{U}(t))\left({\rm E}\{Z_1\} - {\rm E}\{Z_2\}\right)^2,
  \end{split}
\end{align}

\noindent where the isotopic mean of the composite diet is averaged over both seasonal diets, weighted by the proportional inclusion of each.
In the wet/dry season example, the consumer could either shift gradually from its wet season diet to its dry season diet if $\mathcal{U}(t)$ is smooth, or shift abruptly if $\mathcal{U}(t)$ is discontinuous.

\begin{figure}[h!]
\begin{center}
\includegraphics[width=0.6\textwidth]{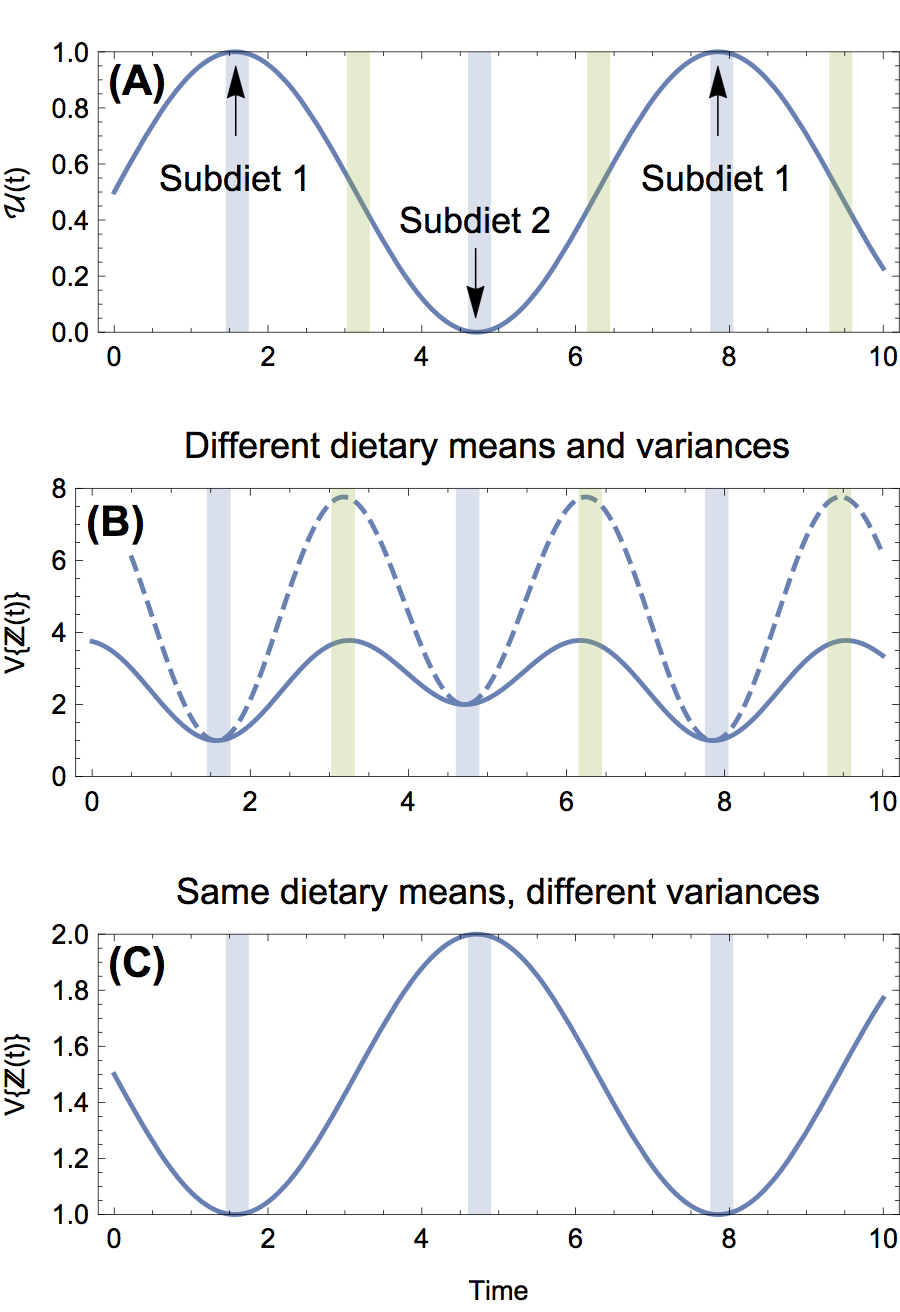}
\end{center}
\textbf{\refstepcounter{figure}\label{figvarZt} Figure \arabic{figure}.}{
{\bf(A)} The proportional contribution of Diet 1 to the composite diet $\mathbb{Z}$ over time.
{\bf(B)} The isotopic variance of the composite diet ${\rm V}\{\mathbb{Z}(t)\}$ when subdiets 1 and 2 have different means and variances. 
Two examples are shown: one where the dietary means are closer in isotopic space (solid), and one where the dietary means are farther apart (dashed).
For both, the peak variance occurs during the dietary transitions (green shading), whereas the troughs reflect the variances of subdiet 1 and 2, respectively (blue shading).
{\bf(C)} The isotopic variance of the composite diet ${\rm V}\{\mathbb{Z}(t)\}$ when subdiets 1 and 2 have the same means but different variances. When the subdiets have the same means, ${\rm V}\{\mathbb{Z}(t)\}$ oscillates to reflect the respective variances of the subdiets, and does not exhibit peak variance during the dietary transition.
}
\end{figure}

Dietary transitions between seasons tend to be gradual, even if the beginning/end of a given season is abrupt \cite{Thompson:1995co,CodronSAJWR2007}.
To understand how a temporally oscillating diet affects the isotopic variance of the composite diet, we consider the smooth oscillation $\mathcal{U}(t) = 1/2 + 1/2\sin(\omega t)$, such that the proportional contribution of $Z_1$ oscillates with frequency $\omega$ (Fig. \ref{figvarZt}A).
Substituting $\mathcal{U}(t)$ into Eq. \ref{eqZtgen} provides the solution to a sinusoidally varying diet, with expectation and variance

\begin{align}
  \begin{split}
  \label{eqZtsin}
  	{\rm E}\{\mathbb{Z}(t)\} &= \frac{{\rm E}\{Z_1\} + {\rm E}\{Z_2\}}{2} + \frac{{\rm E}\{Z_1\} - {\rm E}\{Z_2\}}{2}\sin(\omega t), \\
    {\rm V}\{\mathbb{Z}(t)\} &= \overbrace{\frac{{\rm V}\{Z_1\} + {\rm V}\{Z_2\}}{2} + \frac{1}{2} \left(\frac{{\rm E}\{Z_1\} - {\rm E}\{Z_2\}}{2}\right)^2}^{\alpha_{\rm V}} \\
    &+ \overbrace{\frac{{\rm V}\{Z_1\} - {\rm V}\{Z_2\}}{2}}^{\beta_{\rm V}}\sin(\omega t) + \overbrace{\frac{1}{2}\left(\frac{{\rm E}\{Z_1\} - {\rm E}\{Z_2\}}{2}\right)^2}^{\gamma_{\rm V}}\sin\left(2\omega t + \frac{\pi}{2}\right),
  \end{split}
\end{align}

\noindent where we have combined the non-oscillating components of the variance into three parameters $\alpha_{\rm V}$, $\beta_{\rm V}$, and $\gamma_{\rm V}$ for notational efficiency.

\begin{figure}[h!]
\begin{center}
\includegraphics[width=0.6\textwidth]{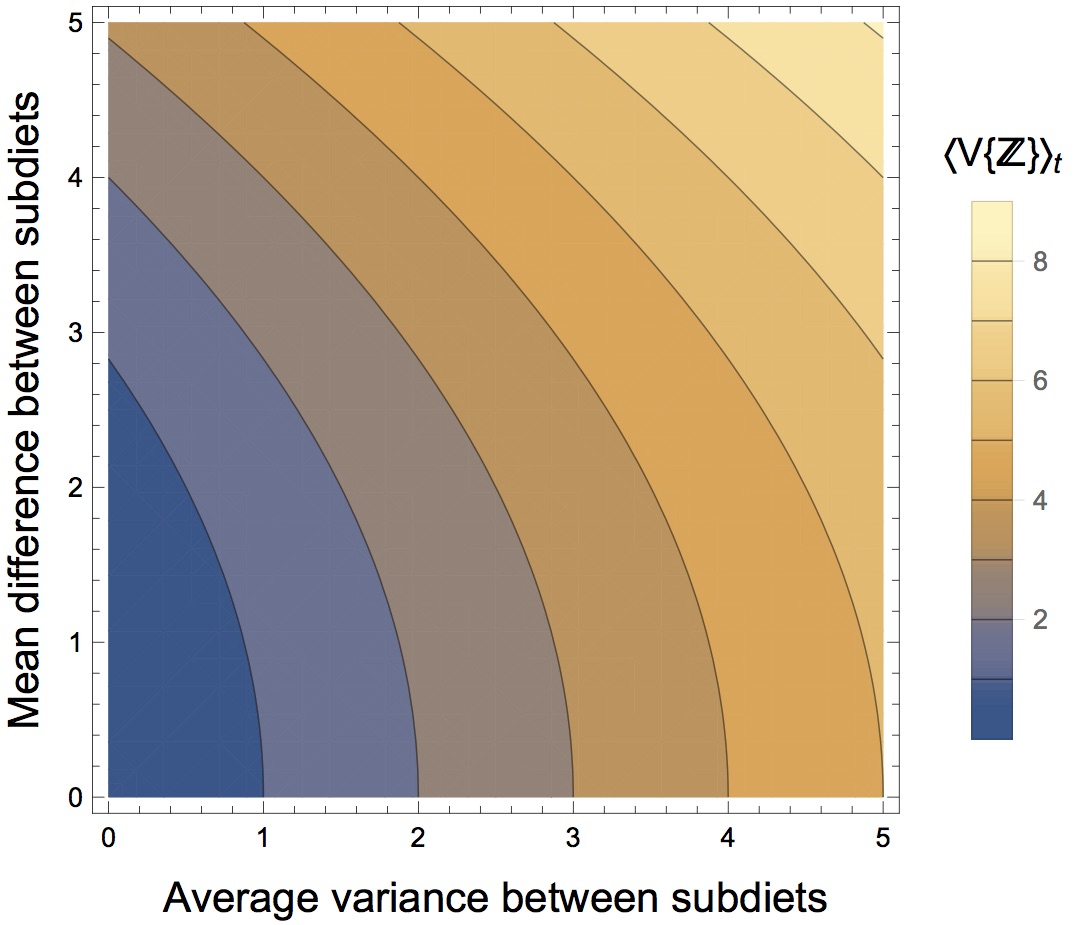}
\end{center}
\textbf{\refstepcounter{figure}\label{figvarcont} Figure \arabic{figure}.}{
Components of the mixing space that affects the time-averaged variance of the composite diet, $\langle {\rm V}\{\mathbb{Z}\} \rangle_t$.
As the average variance between the subdiets increases, the time-averaged variance of the composite diet increases.
As the difference in the isotopic means of the subdiets increase, the time-averaged variance of the composite diet increases, though at a slower rate.
}
\end{figure}

We gain three key insights from the solution for the expectation and variance of the composite diet.
1) As would be expected, the central tendency of the composite diet is the average of the mean values for each subdiet, while the amplitude of oscillations is driven entirely by the difference in the mean values of each subdiet;
2) the time-averaged variance (denoted by $\langle \cdot \rangle_t$) is simply $\langle {\rm V}\{\mathbb{Z}(t)\} \rangle_t = \alpha_{\rm V}$, is only impacted by the average variance between the seasonal diets and the difference in the mean isotope values between the seasonal diets (Fig. \ref{figvarcont});
3) the oscillating component shows that the composite dietary variance has a modified frequency, as well as an offset, meaning that the maximal variance of the consumer's composite diet generally occurs during the transition from one diet to the other (Fig. \ref{figvarZt}B).
Together, these results reveal that if the consumer's diet is varying continuously between two seasonal diets over time, both the averaged variance, as well as the difference in the mean isotope values of the seasonal diets -- directly reflecting the heterogeneity of prey mixing space geometry -- will serve to increase the variance of the consumer's diet averaged over time, and by extension the isotopic variance of the consumer itself.

We also observe that the consumer's peak variance (its maximum niche width) occurs not during the exclusive adoption of either subdiet, but during the transition between the two, and that the magnitude of the peak variance is driven exclusively by the difference in isotopic means between seasonal diets.
As the seasonal diets become more heterogeneous in isotopic space, the greater the consumer's peak variance during the transition, and this occurs because it is sampling between two dietary strategies that are isotopically distinct.
We can directly observe this by considering a transition between two diets with  a) different means and equal variances, and b) equal means and different variances.
In the former case, the peak variance of the composite diet occurs during the transition with magnitudes determined by the difference in isotopic means between subdiets (Fig \ref{figvarZt}B); in the latter case, because the diets have the same mean isotope value, the peak occurs not during the transition, but when the consumer adopts the diet with greater variance, which in our example would occur at the height of the season (Fig \ref{figvarZt}C).

\begin{figure}[h!]
\begin{center}
\includegraphics[width=0.75\textwidth]{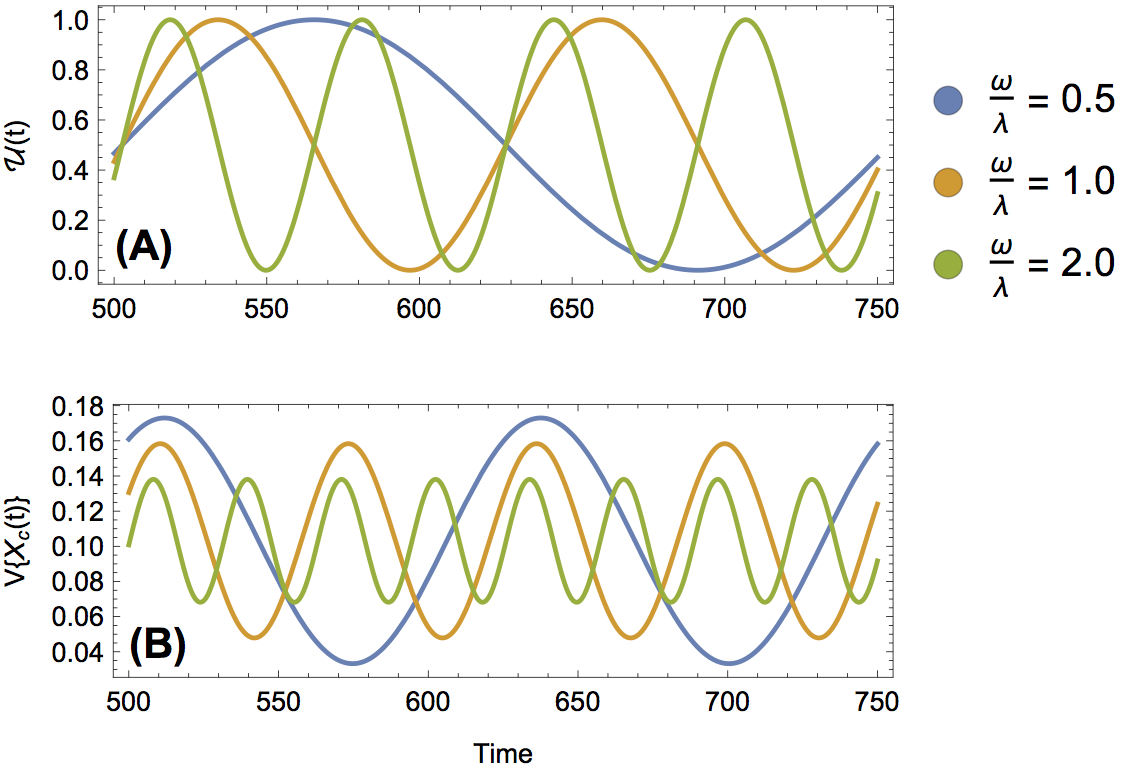}
\end{center}
\textbf{\refstepcounter{figure}\label{figxcsin} Figure \arabic{figure}.}{
{\bf(A)} A sinusoidally varying diet, where $\mathcal{U}(t) = 1/2 + 1/2\sin(\omega t$) for increasing values of $\omega$, or equivalently, decreasing timescales of dietary switching.
{\bf(B)} The isotopic variance of a consumer over time ${\rm V}\{X_c(t)\}$ across increasing values of $\omega$ relative to the consumer's incorporation rate $\lambda$. 
As the timescale of diet switching decreases relative to the timescale of isotopic incorporation, the amplitude of isotopic variance decreases due to increased isotopic averaging over faster shifts in diet.
}
\end{figure}

The isotopic composition of a consumer $X_c(t)$ during a single dietary shift is governed by a single timescale of \emph{physiological} origin: the rate of incorporation $\lambda$ (Eq. \ref{eqEVar}).
However, a seasonally shifting diet that is driven by oscillating foraging strategies introduces an additional timescale of \emph{ecological} origin that will affect $X_c(t)$, determined by the frequency of diet switching $\omega$ (Fig. \ref{figxcsin}A).
Depending on the turnover rate of the tissue of interest and how often the consumer shifts its diet, the ratio of these timescales $\omega/\lambda$ will impact how the isotopic mean and variance of the consumer changes over time.
For the case of a sinusoidally varying diet, we can solve for $X_c(t)$ directly, such that

\begin{align}
  \label{eqXsin}
  \begin{split}
    {\rm V}\{X_c(t \to \infty)\} = &\alpha_{\rm V}\frac{\lambda}{2} + \beta_{\rm V}\frac{\lambda^2}{\sqrt{(2\lambda)^2 + \omega^2}}\sin\left(\omega t -\theta_1\right) \\
    &+ \gamma_{\rm V}\frac{\lambda^2}{2\sqrt{\lambda^2 + \omega^2}}\sin\left(2\omega t + \theta_2\right).
  \end{split}
\end{align}

\noindent where the offsets $\theta_1$ and $\theta_2$ are $ \tan^{-1}(\omega/2\lambda)$ and $ \tan^{-1}(\lambda/\omega)$, respectively.
As in the case of a single diet $Z$, the time-averaged variance is scaled by the incorporation rate as $\alpha_{\rm V}\lambda/2$.
Moreover, we observe that the consumer's isotopic composition lags behind changes in diet, such that an isotopic shift in the consumer's tissues is observed after the actual foraging shift.
This lag involves both $\theta_1$ and $\theta_2$, however these offsets play different roles in contributing to the lag for different mixing space geometries.
When the isotopic means of the seasonal diets are similar, the lag is mostly due to $\theta_1$; when the means are different and the variances are similar, the lag is mostly due to $\theta_2$; when both the isotopic means and variances of the seasonal diets are different, both contribute significantly to the lag.

As shown in Fig. \ref{figxcsin}, we observe that
1) the lag between the transition and the peak variance of the consumer increases with decreasing $\lambda$ (i.e. increasing timescale of incorporation), and
2) the amplitude of the variance of $X_c(t)$ decreases with increasing $\omega$ (i.e. decreasing timescale of ecological switching).
The first result is not surprising, as it mirrors the role of $\lambda$ in the static diet example. 
The second result is less intuitive: in words, as the consumer shifts its diet more frequently, there is still a peak variance during dietary transitions, though with diminishing amplitude, and this would make it more difficult to measure (Fig \ref{figxcsin}B).
The decrease in the amplitude of isotopic variance of the consumer's tissue is thus an averaging effect, where the timescale of incorporation is much larger than the timescale of dietary switching.

\section{Discussion}


We have established a forward-integration approach towards understanding how the isotopic distribution of an individual consumer evolves due to ecological, physiological, and geometric factors.
Our framework introduces mechanistic links between the ecological foraging dynamics of a consumer, the physiological constraints that dictate incorporation, and the more abstract effects of mixing space geometry, such as the heterogeneity of prey isotope distributions.
We focus our efforts on building an analytical framework to understand how the isotopic variance of an individual -- its isotopic niche width -- changes as a function of different foraging strategies that are both probabilistic and dynamic over time.
We consider two foraging scenarios:
1) {\it static strategy}: probabilistic consumption of multiple prey, the proportions of which are on average constant over time, and
2) {\it dynamic strategy}: probabilistic consumption of multiple prey whose relative contribution to the consumer's diet varies over time.

Our primary findings concern whether and to what extent the peak isotopic variance of the consumer, or maximum isotopic niche width, is realized under different, but definable, conditions with respect to static and dynamic foraging scenarios.
When the consumer exhibits a static foraging strategy, the isotopic variance of its diet is tied directly to its prey specialization and the skewness of the isotopic mixing space.
We show that as the mixing space becomes more skewed, there is an increasing likelihood that the peak variance will occur at intermediate specialization (where a single prey accounts for ca. 50\% of the consumer's diet; Figs. \ref{figvar},\ref{figspecvar}).
When the consumer exhibits a dynamic, yet smoothly varying foraging strategy, we show that the peak variance occurs during the transition from one diet to another, and is offset by a lag that is a function of both its incorporation rate and the timescale over which it shifts between diets (Fig. \ref{figvarZt}).
Below we show that these findings have relevance by examining an empirical sea otter mixing space, and discuss areas where additional realism can be incorporated to gain further ecological insight into the isotopic niche.

\subsection*{The isotopic niche: generalization vs. specialization}

\begin{figure}[h!]
\begin{center}
\includegraphics[width=0.75\textwidth]{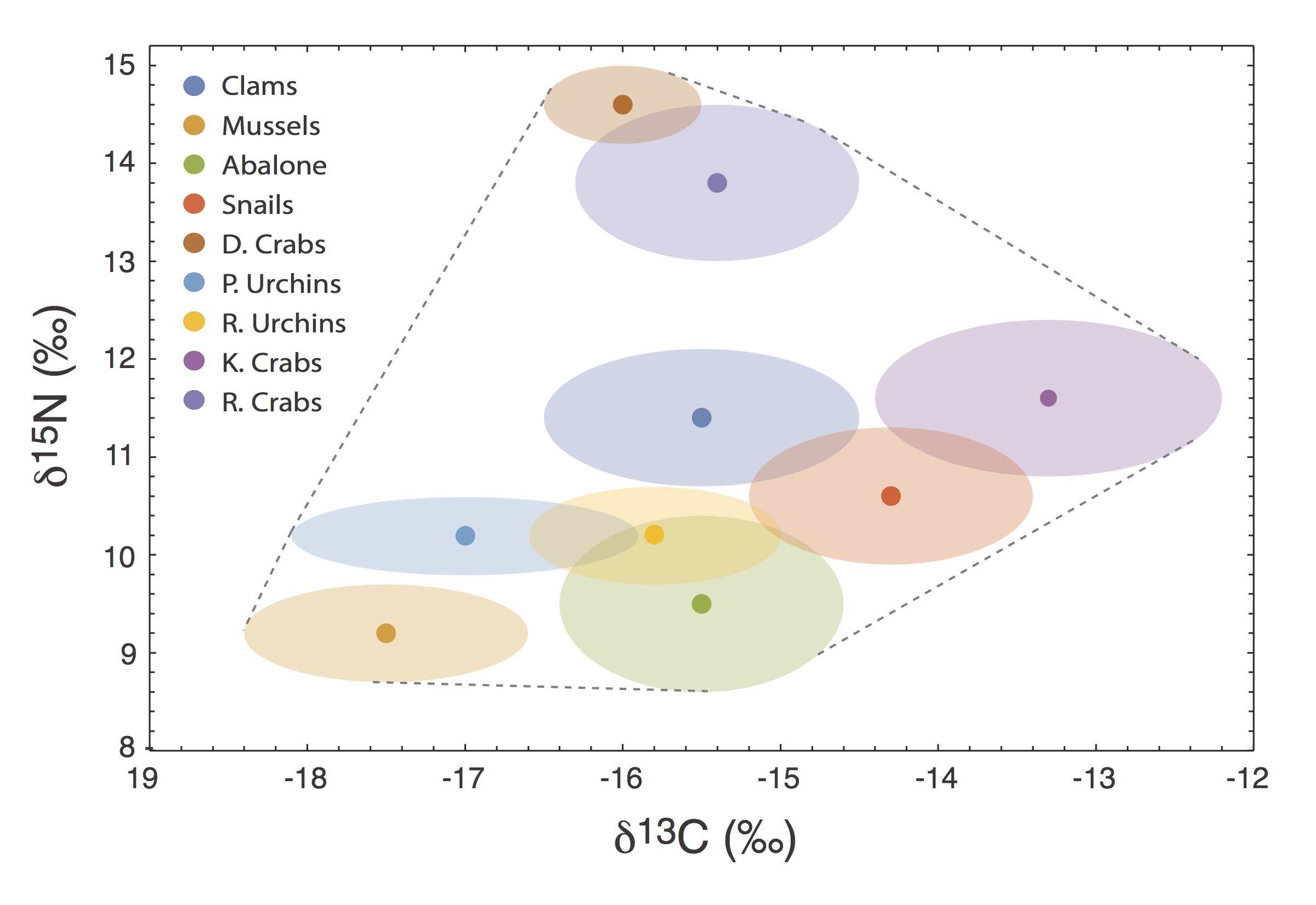}
\end{center}
\textbf{\refstepcounter{figure}\label{figsomix} Figure \arabic{figure}.}{
The isotopic mixing space ($\delta^{13}{\rm C}$ vs. $\delta^{15}{\rm N}$) for a sea otter consumer near San Simeon and Monterey Bay, California, composed of nine commonly consumed invertebrate species. Units are per-mil ($\permil$).
}
\end{figure}
\begin{figure}[h!]
\begin{center}
\includegraphics[width=1\textwidth]{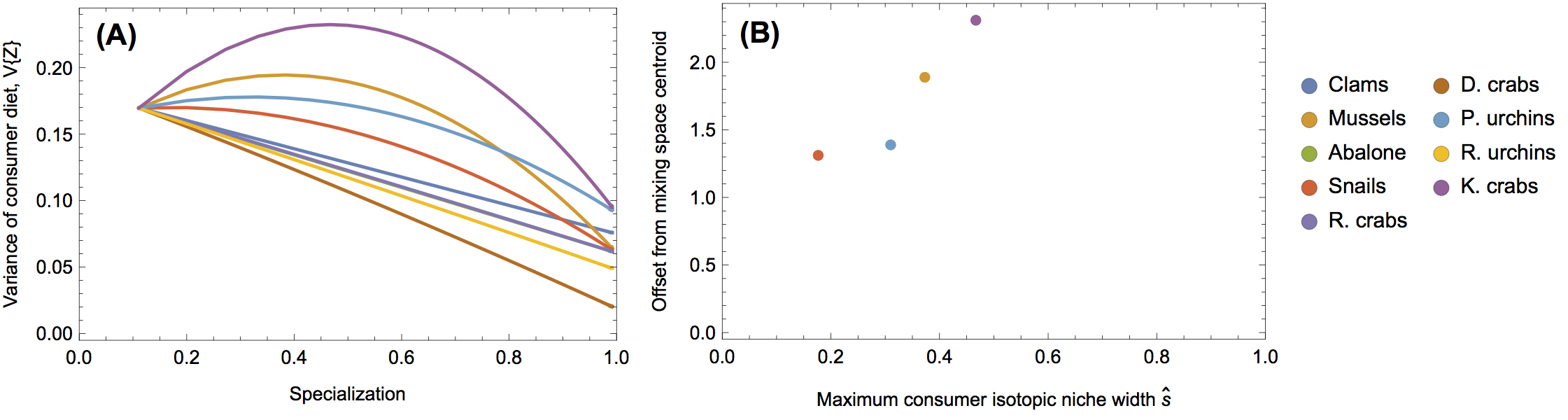}
\end{center}
\textbf{\refstepcounter{figure}\label{figottervar} Figure \arabic{figure}.}{
{\bf(A)} Predicted variance in $\delta^{13}{\rm C}$ of sea otter diets over different degrees of specialization on each prey in the system (colors).
{\bf(B)} Calculated maximum consumer niche width values as a function of specialization and the offset of the prey isotopic mean from the mixing space centroid.
}
\end{figure}

To demonstrate the empirical relevance of the nonlinear nature of ${\rm V}\{Z\}$, we examine a prey-rich marine system near San Simeon and Monterey Bay, California, composed of nine invertebrate species commonly consumed by sea otters \cite{Tinker:2008tp}.
In this system, all potential prey resources have unique isotopic means and variances (Fig. \ref{figsomix}), including multiple species of sea urchins and crab, clams, abalone, mussels, and snails.
We can investigate how alternatively targeting each prey species alters the isotopic variance of a sea otter's diet across different degrees of specialization by modifying the underlying Dirichlet distribution (i.e. by increasing $a_k$ for each species individually, while holding $a_{i \neq k} = 1$; Fig. \ref{figottervar}A).
We determined the existence of strong nonlinearity in the isotopic variance of diet for 44\% of prey species (Fig. \ref{figottervar}B). 
For targeted prey exhibiting nonlinear variance (including mussels, snails, purple sea urchins, and kelp crabs), the maximum isotopic variance was found in the region $s \leq 0.5$.

The sea otter example reveals that the parabolic nature of the isotopic variance of a consumer's diet predicted by our statistical model has particular relevance for real-world prey mixing space geometries.
The message is straightforward: for a given prey mixing space, a consumer's dietary variability -- where the consumer's tissues scale in proportion to its diet by a factor of $\lambda/2$ -- will be a function of both mixing space geometry, as well as its dietary strategy, and these effects can be confounding.
Despite this, we are able to establish certain predictions for the consumer's isotopic niche width as a function of diet: as the consumer incorporates moderate amounts of isotopically unique prey into its diet, its variance will be expected to increase.
Knowledge of the interplay between mixing space geometry and a consumer's dietary strategy, and its consequent effect on the isotopic variance of diet, is particularly important for characterizing consumers based exclusively on isotopic variance.
For example, without knowledge of these relationships, a highly variable consumer might be interpreted as a dietary generalist, whereas it might be able to achieve a similarly high or higher variance by moderately specializing on a single prey species with an isotopic distribution far from the mixing space centroid.

\subsection*{The isotopic niche over time}

We gain additional insight into the factors influencing consumer isotopic variability by considering dynamic diets, where the consumer oscillates between different foraging strategies over time.
We considered a simple sinusoidal oscillation for $\mathcal{U}(t)$, the proportional contribution of subdiet 1 to the composite diet over time $\mathbb{Z}(t)$.
The subdiets from which the composite diet is composed can be thought of as `seasonal diets'.
Our analytical results showed that the peak variance of the composite diet occurred during the transition between seasonal diets.
Importantly, this is not due to any particular mixing space geometry, but a general result that will always occur, as long as the diets are isotopically distinct (each with a unique mean and variance), and the transition is smooth such that a consumer gradually shifts between different diets, as opposed to an abrupt, discontinuous diet switch.

Although the peak variance of $\mathbb{Z}(t)$ is entirely due to ecological diet shifts rather than mixing space geometry, the latter does play a role in determining the mean value (time average) of ${\rm V}\{\mathbb{Z}(t)\}$.
The effect of mixing space geometry on the time-averaged variance of the composite diet is determined by $\alpha_{\rm V}$, which is a function of
1) the average variance of the subdiets from which $\mathbb{Z}$ is composed, and
2) the mean difference between the two subdiets (see Eq. \ref{eqXsin}).
As either of these factors increase, the average variance of the composite diet increases, setting the baseline from which the peak dietary variance fluctuates (Fig. \ref{figvarcont}).
We also observed that as the frequency of dietary transitions increased relative to the rate at which the consumer integrates dietary isotopes into its tissues, the consumer's isotopic variance exhibited lower amplitude as it fluctuated between the different variances of its diet (Fig. \ref{figxcsin}).
This occurs because the greater transition frequencies serve to average variance of the two diets within its tissues.
\cite{Fink:2012eg} found a similar dynamic when they derived an analytical solution for the variance of a consumer population transitioning between two prey.

An interesting observation that we gain from exploring a sinusoidal dietary shift is that the variance peak observed during dietary transitions is dependent on the smoothness of the transition.
In fact, it is the transition mid-point, at $\mathcal{U}(t) = 0.5$, where the composite diet is pulled equally from each seasonal diet, and this serves to maximize the isotopic heterogeneity of the mixture (the consumer).
Thus, when foraging strategies are dynamic, it is the point of maximum isotopic heterogeneity that results in peak isotopic variance of the diet.
This is analogous to the cause of peak dietary variance in the static example, where specialization on prey resources with greater isotopic offsets from the mixing space centroid maximizes isotopic heterogeneity, resulting in a variance peak.

\begin{figure}[h!]
\begin{center}
\includegraphics[width=0.75\textwidth]{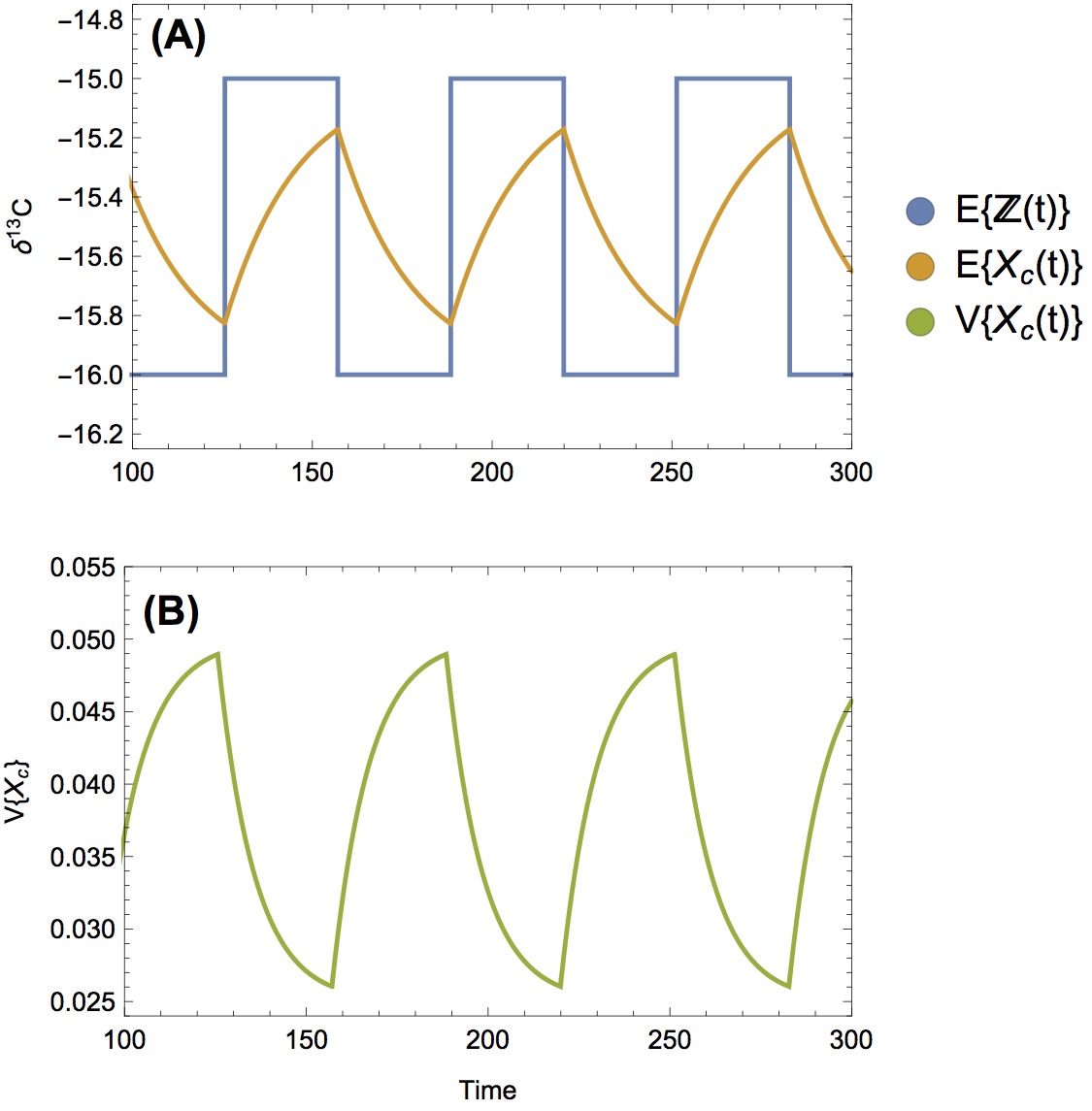}
\end{center}
\textbf{\refstepcounter{figure}\label{figsawtooth} Figure \arabic{figure}.}{
{\bf(A)} The mean isotopic value of the composite diet over time ${\rm E}\{\mathbb{Z}(t)\}$ when diet-switching is discontinuous, following a square wave pattern, where subdiets have a mean $\delta^{13}{\rm C}$ value of $-15$ and $-16$, respectively.
The mean isotopic value of the consumer over time ${\rm E}\{X_c(t)\}$ is observed to abruptly change directions when its diet transitions, asymptoting towards (but not reaching) the isotopic mean of its current diet.
{\bf(B)} Consumer isotopic variance ${\rm V}\{X_c(t)\}$ follows a similar trajectory over time, asymptoting towards (but not reaching) the isotope variance of its current diet.
When diets follow a discontinuous switching dynamic, the peak variance does not appear at the transition, as it does when the diet switching is smooth.
}
\end{figure}

An extreme alternative to a smooth dietary transition would be one that is discontinuous, as depicted by a step-function, or square wave (Fig. \ref{figsawtooth}A).
Such an instantaneous dietary shift is not ecologically unrealistic; e.g. both brown bears and gray wolves abruptly shift their diet to salmon during salmon runs \cite{Hilderbrand:1999kq,Darimont:2002ex,Levi:2012ir}, as do predators on other prey populations exhibiting localized boom-bust dynamics such as locusts, krill, jellyfish, and sardines \cite{Dawson:2008hj,Atkinson:2014jm}.
Because there is no point during a sharp, discontinuous transition that serves to mix subdiets, the variance of the composite diet does not peak in response.
Instead, both the expectation and the variance of the composite diet incorporates this step function behavior, transitioning to reflect the shifts between different diets.
Because the isotope ratios associated with diet are incorporated gradually into the consumer's tissues, both the mean and the variance of the consumer will adopt a sawtooth-like dynamic (Fig. \ref{figsawtooth}A,B), where they begin to asymptote to the expectation and variance of the subdiets, but are reverted abruptly at the dietary switch.
As in the static example, the isotopic variance of the consumer approaches the variance of its diet twice as fast as its expectation (cf. Eq. \ref{eqEVar}).

\subsection*{Population dynamics and state-dependent foraging}
One potentially important extension of our framework could incorporate a population dynamic underlying the availability of potential resources (and by extension the consumer's diet) in a continuous, more complex, and ecologically justified manner.
Our original formulation of the Dirichlet distribution that describes the consumer's diet was established on the relationship between the random variables describing the proportional contribution of prey to diet ($P_i = p_i$) and its encounter rate ($\Psi_i = \psi_i$), where $p_i = \psi_i/\sum_j \psi_j$, and this was assumed to have a static distribution over time.
However, if the prey are fluctuating in accordance to an underlying population dynamic (for example, determined by a system of differential equations), the encounter rate of each prey would itself be a function of time.
By relating the expectation and/or variance of $\Psi_i$ to the density of prey, the parameterization of the Dirichlet can be directly coupled to changes in population densities, thus mechanistically incorporating population dynamics into predictions of a consumer's isotopic composition.

Furthermore, the relationship between $p_i$ and $\psi_i$ explicitly assumes passive foraging between the consumer and its potential prey, and this holds for our original static (single diet) example, our shifting diet example, and would hold for the example above where the Dirichlet changes in response to an underlying population dynamic.
Although this is not a bad starting point, and may be a perfectly reasonable assumption for a filter feeder that consumes resources indiscriminately, it is not a reliable assumption for most organisms that may rank prey based on intrinsic traits (e.g., energetic yield, handling/processing time).
Instead, a more complex relationship between $p_i$, the traits of the consumer's prey, and perhaps traits of the consumer itself, could be used to determine the parameterization of the Dirichlet distribution defining the consumer's diet over time.

For example, our framework implicitly assumes that there is a steady state influx of prey biomass to match the metabolic expenses of the consumer.
In other words, prey are chosen in accordance to the Dirichlet distribution, but it is assumed that each foraging bout contributes equally to the consumer's diet, and that the consumer always finds a meal.
In reality, the success of a given foraging bout is not certain, and there is some risk of not finding any prey at all \cite{Creel:2008p838}; in such a case, the consumer would resort to metabolizing its own tissues \cite{Doucett:1999bz,VanderZanden:2001p2449}.
Such a dynamic would directly impact the rate of incorporation by altering the proportional contribution of newly consumed isotopes to the turnover of the consumer's body tissues.
Moreover, the foraging decisions that a consumer makes are often a function of its energetic state \cite{Barnett:2007er,Yeakel:2013hi}, which changes as it successfully or unsuccessfully finds and acquires its prey \cite{Mangel:1986um}.
Such state-dependent foraging may be difficult to treat analytically, but could be explored numerically, and this approach would be useful for hypothesis testing, particularly when one is interested in comparing the effects of different foraging strategies on the statistical properties of the consumer's isotopic composition.


\subsection*{From individual consumers to populations}
Finally, the framework that we have presented has focused entirely on the individual, in particular on how the isotopic variance of an individual consumer changes in response to different ecological and physiological factors as well as aspects of the isotope mixing space it utilizes.
Most ecological applications using stable isotope analysis operate at the level of the population, although there is a rich history of using stable isotopes to understand sources of dietary variation at the level of the individual \cite{Koch:1995vj,Matthews:2004hw,Sponheimer:2006fj,Post:2008ki,Newsome:2009tn,Yeakel:2009hz,Newsome:WhhVfocb,Hopkins:2015ip}.

\begin{figure}[h!]
\begin{center}
\includegraphics[width=0.75\textwidth]{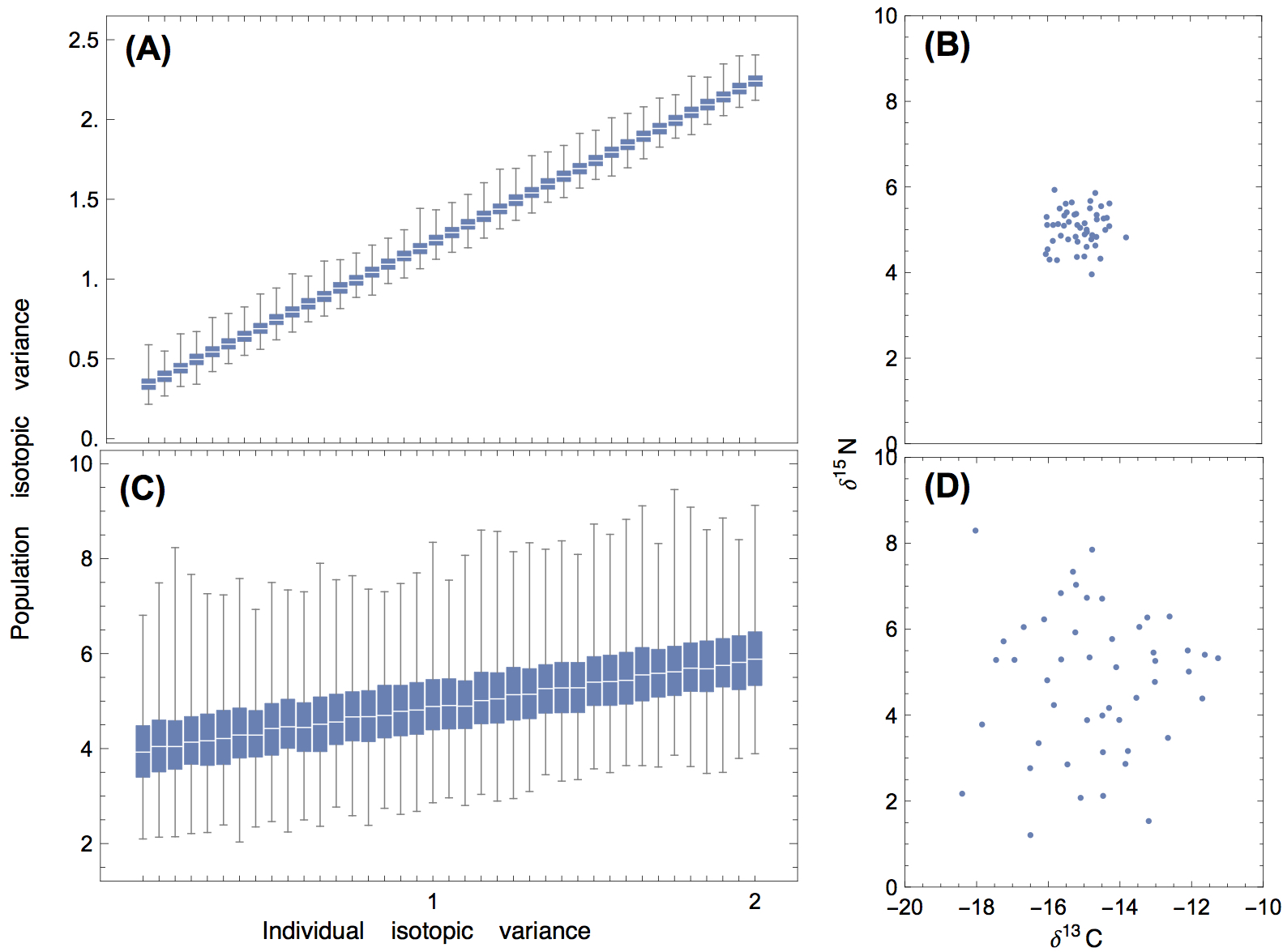}
\end{center}
\textbf{\refstepcounter{figure}\label{figindpopvar} Figure \arabic{figure}.}{
Population isotopic variance vs. individual isotopic variance where mean isotope values of individuals are randomly drawn from a Normal distribution (50 individuals per population; 1000 replicates), and individuals are assumed to have the same variance.
{\bf(A)} and {\bf(B)} When the mean values of individuals are randomly drawn from a normal distribution with low variance, there is a linear relationship between individual-level and population-level isotopic variance.
{\bf(C)} and {\bf(D)} When the mean values of individuals are randomly drawn from a normal distribution with high variance, the relationship becomes masked by noise.
}
\end{figure}

Understanding how variance percolates from prey to the individual consumer is a necessary first step for understanding sources of isotopic variation at the level of the population.
This is not always straightforward, as the isotopic variance of an individual may or may not be closely coupled with the variance of the population.
For example, if individuals within a population have similar means and - for simplicity - equal variances, then the variance of the population will scale linearly with the variance of the individuals (Fig. \ref{figindpopvar}A,B).
This relationship highlights an important message: when individuality is low, the variance of the population is entirely explained by the variance of the individuals; this means that the results that we have presented for a consumer individual are expected to scale directly to that of the population. 
However, if the individuals within a population have very different means and relatively small variances, then there will not be a significant relationship between population-level and individual-level variation (Fig. \ref{figindpopvar}C,D).

We can imagine different individual-population relationships occurring within a 2-D state-space defined by individuality on one axis and specialization on the other.
At the extremes, a population could consist of
1) obligate specialists with low individuality where all individuals specialize on the same resource,
2) obligate specialists with high individuality where all individuals specialize on different resources, and
3) obligate generalists with low individuality where all individuals are generalists; an obligate generalist with high individuality cannot exist in this context. 
These potential end-members are discussed at length in \cite{Bearhop:2004im} and \cite{Fink:2012eg}.
As we have seen in the above analyses, the isotopic variance of individuals is driven by an interplay between mixing space geometry, consumer foraging behaviors, and physiological incorporation.
How these different population-level end-members might shape both individual and population-level isotopic distributions is an important question, though the answers will likely harbor additional complexities.
For example, isotopically similar individuals with low variances imply that all individuals are consuming similar things, in similar quantities, such that individuality is low, though our results show that low isotopic variance need not indicate specialization or generalization \emph{per se} (cf. Figs \ref{figvar},\ref{figottervar}).
Accounting for individual variation in dietary proclivities over time is bound to complicate interpretation further.


%

\subsection*{Conclusions}

There are many sources of variation that contribute to a consumer's isotopic composition.
These sources include the geometry of the prey mixing space, the foraging behaviors of the consumers, as well as temporal changes in the environment that might alter the ability of the consumer to find, acquire, and consume its prey.
Along with physiological incorporation of isotopes into consumer tissues, these factors serve to drive the temporal evolution of the consumer's isotopic distribution, or isotopic niche.
By coupling the isotopic variance of this distribution to mechanistic relationships between the consumer and its diet, as well as the isotopic mixing space of the system, we have presented a systematic exploration of the factors that cause the isotopic niche to both expand and contract.
Incorporating the effects of population dynamics and/or more realistic foraging strategies will enable hypothesis testing of different ecological mechanisms to generate the isotopic distributions that are observed in nature.
We hope that such a forward-integrating approach, alongside the use of tools such as mixing models to back-calculate dietary composition, will serve to expand and enhance the ecological interpretation of isotopic data.

\section*{Disclosure/Conflict-of-Interest Statement}

The authors declare that the research was conducted in the absence of any commercial or financial relationships that could be construed as a potential conflict of interest.

\section*{Author Contributions}
JDY, UB, EAES, and SDN conceived the idea. 
JDY and UB designed the statistical framework and conducted the analyses. 
EAES and SDN contributed empirical data. 
All authors contributed equally to drafting and writing the manuscript. 

\section*{Acknowledgments}
We thank C.E. Chow, J.P. Gilbert, A. Jakle, M. Mangel, Z.D. Sharp, F.A. Smith, and the Center for Stable Isotopes Brown Bag Symposium at the University of New Mexico for helpful comments and discussions that greatly increased the quality of this manuscript.

\section*{Funding} J.D. Yeakel was supported by the Omidyar Postdoctoral Fellowship from the Santa Fe Institute. U. Bhat was funded in part by grant No. 2012145 from the United States-Israel Binational Science Foundation (BSF) and Grant No. DMR-1205797 from the National Science Foundation.


\newpage
\section*{References}


\newpage

\begin{table}[!t]
\textbf{\refstepcounter{table}\label{Tab:02} Table \arabic{table}.}{ Parameters and their definitions.}

\begin{tabular}{ll}
\hline
Parameter & Definition\\ 
\hline
$n$ & Number of potential prey in the consumer's diet\\
$M_i = m_i$ & Number of encounters of prey $i$\\
$\Psi_i = \psi_i$ & Encounter rate of prey $i$\\
$a_i$ & Dispersion of prey $i$ ($\propto$ encounter rate)\\
$c$ & Scales with time between prey encounters\\
$P_i = p_i$ & Proportional contribution of prey $i$ to diet\\
$Z = z$ & Isotopic composition of the consumer's diet\\
$\mu_i$ & Isotopic mean of prey $i$\\
$\sigma^2_i$ & Isotopic variance of prey $i$\\
$s_k$ & Specialization of the consumer on prey $k$\\
$\lambda$ & Rate of isotopic incorporation\\
$X_c(t)$ & Isotopic composition of the consumer over time\\
${\mathbb Z}(t) = \mathbb{z}(t)$ & Isotopic value of the composite diet over time\\
${\mathcal U}(t)$ & Proportional contribution of subdiet $Z_1$ over time\\
$\omega$ & Frequency of diet switching\\
\end{tabular} 
\end{table}

\newpage

\end{document}